\documentclass[showpacs,floatfix,prd,twocolumn]{revtex4-1}

\usepackage[T1]{fontenc}
\usepackage[utf8x]{inputenc}
\usepackage{amsmath}
\usepackage{amsfonts}
\usepackage{amssymb}
\usepackage{gensymb}
\usepackage{mathrsfs}
\usepackage{latexsym}
\usepackage{graphicx}
\usepackage{nicefrac}
\usepackage[ps2pdf]{hyperref}

\hypersetup{
  pdfauthor={C. E. Pellicer, Elisa G. M. Ferreira, Daniel C. Guariento,
    André A. Costa, Leila L. Graef, Andrea Coelho, Elcio Abdalla}
  pdftitle={The role of dark matter interaction in galaxy clusters},
}

\begin{document}

\title{The role of dark matter interaction in galaxy clusters}



\author{C. E. Pellicer}
\email{carlosep@fma.if.usp.br} 

\author{Elisa G. M. Ferreira}
\email{elisa@fma.if.usp.br}

\author{Daniel C. Guariento}
\email{carrasco@fma.if.usp.br}

\author{\mbox{Andr\'{e} A. Costa}}
\email{alencar@fma.if.usp.br}

\author{Leila L. Graef}
\email{leila@fma.if.usp.br}

\author{Andrea Coelho}
\email{andreac@fma.if.usp.br}

\author{Elcio Abdalla}
\email{eabdalla@fma.if.usp.br}

\affiliation{Instituto de F\'{\i}sica, Universidade de S\~ao Paulo,\\
CP 66318, 05315-970, S\~ao Paulo, Brazil}


\begin{abstract}

We consider a toy model to analyze the consequences of dark matter interaction with a dark energy background on the overall rotation of galaxy
clusters and the misalignment between their dark matter and baryon distributions when compared to $\Lambda$CDM predictions. The interaction
parameters are found via a genetic algorithm search. The results obtained suggest that interaction is a basic phenomenon whose effects are detectable
even in simple models of galactic dynamics.
\end{abstract}


\pacs{98.65.Cw, 98.80.Es, 95.35.+d, 95.10.Ce}

\maketitle

\section{Introduction}

Today, it is basically taken for granted that as much as one quarter to one third of the energy content of the Universe is composed by a strange and
heavy kind of matter, the so called dark matter. An even stranger negative pressure object, responsible for its acceleration, the so called dark
energy, fills about two thirds. Baryons are responsible for less than five percent of the energy content of the Universe
\cite{perlmutter-1998,*bertone-2005}.

In general one assumes that dark energy is described by a cosmological constant, an assumption compatible with the recent WMAP
data\cite{komatsu-2011,*larson-2011}. However, there is a cloud upon these results, since the theoretical expectation for a cosmological constant, if
not vanishing by some unknown and unexpected symmetry, is 120 orders of magnitude above the observational value. Moreover, there is a further question
mark concerning why the dark energy is important exactly today, or equivalently why its value coincides with today's energy content of the
Universe. This question led several authors to propose a model for dark energy, which naturally interacts with dark matter
\cite{zimdahl-2001,*chimento-2003,*rosenfeld-2007,*quartin-2008,*wu-2008,*setare-2008,*gumjudpai-2005,*wang-2005,*wang-jcap-2008,*feng-2008,*wang-2006,*wang-2007,*wang-2008}.

It has also been realized that galaxy clusters may contain information about dark energy and dark matter interaction because the hidden sector
interaction implies a correction to the virialization process in the cluster\cite{bertolami-2007,*bertolami-2009}, leading to
quite strong constraints in the interaction parameters \cite{abdalla-2009,*abdalla-2010,*he-2010}.

More recently, a displacement in the angular distribution of baryon matter with respect to dark matter in galaxy clusters has been detected through
gravitational lensing \cite{oguri-2010}, and it has been suspected that the alignment of satellite galaxies in clusters is affected by dark matter
interaction \cite{baldi-2011}. In fact, hints to a contrast between the $\Lambda$CDM results and observations in the cluster matter distribution may be
relevant to the overall matter distribution, since Lee \emph{et al.} \cite{lee-2010} observed a departure from the $\Lambda$CDM prediction when
confronted to the fact that the orientations of the galaxy distributions are weakly correlated to the dark matter distribution in the cluster. In
\cite{lee-2010} it has been claimed that such a contrast with the $\Lambda$CDM result is consistent at a 99\% confidence level. As this is a highly
important conclusion, an explanation is mandatory.

Here we follow a similar trend, showing that a displacement in the angular distribution can be traced back to dark matter interaction. Indeed, if dark
matter interacts with dark energy, there will be a kind of external potential for dark matter that does not affect baryon matter and the behavior of
dark matter with respect to baryons will be unbalanced. Such a behavior has already been analysed in the case of galactic clusters in some recent
papers \cite{bertolami-2007,bertolami-2009,abdalla-2009,abdalla-2010} with a positive (sometimes marginal) answer for the interaction. Now we consider
the effect of the interaction on the rotation of clusters.

Our aim is to consider a very simple model for the cluster and show that an external potential mimicking the interaction of the dark sector is enough
to lead to results similar to the observation. We consider the interaction to be, generally speaking, such that its strength is proportional to the
existing matter density, namely

\begin{equation}\label{interacao} 
  \dot{\rho}_{\text{DM}} + 3 H \rho_{\text{DM}} = g H \rho\,,
\end{equation}

\noindent
where $H$ is the Hubble constant, $g$ is a coupling to be determined and $\rho$ in the right-hand side is a combination of dark matter and dark
energy \cite{feng-2008}. Equation \eqref{interacao} gives us a phenomenological effect which might account for the recent observations of the clusters
presenting this anomaly. Such an interaction has been widely used to model the dark sector interaction, although it is not the only possible
form. Indeed, the two-fluid interaction has several different possible forms and can also be originated from a field-theory-inspired interaction,
which we do not pursue here.

\section{A toy model for cluster dynamics}

There are strong evidences in cluster dynamics to encourage the notion that the dark matter concentration has a dynamics of its own and sometimes does
not follow the baryon component. In dark matter interaction, there is no scattering or diffusion, but only gravitational interaction governing the
peculiar movements (mostly rotation). Individual galaxies also interact with each other mainly by gravitation. The classical interaction is sufficient
to provide a means of separating dark matter and galactic baryons. However, it has been pointed out \cite{lee-2010} that a purely gravitational
interaction (or else, a $\Lambda$CDM scenario) does not fit the observations and the dark matter follows a different dynamics. In such a case a
non-gravitational force would be required. Moreover, such a force is highly improbable for baryons, whose interactions are too well known to allow for
any new interaction. There is a quite natural new interaction for the dark sector in case the dark energy is not a cosmological constant, but a new
field, presumably part of the particle physics family, or an extended new standard model. The dark energy interaction with the dark matter implies the
existence of an external potential for the dark matter behavior.

One consequence of this interaction may be observed in astrophysical objects. It is a result from cosmological simulations that galaxy clusters formed
from gravitational collapse are triaxial \cite{frenk-1988}. After virialization, or sufficiently close to it, the cluster evolution may be
characterized by rigid-body parameters, such as a net angular momentum relative to the axes. Thus, a simple model for the dynamics of a cluster as a
whole, disregarding individual components, would consist of a system formed by two overlapping triaxial prolate ellipsoids, corresponding to the dark
and baryon distributions. Their effective radius and angular momentum would depend on their density profiles, and ultimately be determined by
observation and accretion models.

For our purposes, we go one step further on our simplifying hypotheses. We assume that the rotation is confined to only one axis. This assumption is
supported by the fact that large mass clusters have an increased ellipticity \cite{bailin-2005}. Therefore, such a system is sufficiently well described
by a rigid bar or a rigid set of collinear point masses.

\begin{figure}[!htp]
  \centering
  \includegraphics[width=.45\textwidth]{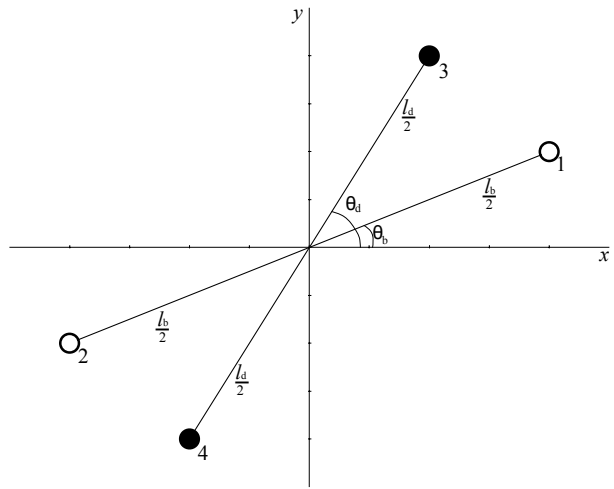}
  \caption{Toy model for cluster matter disposition. The white spheres (1, 2) correspond to baryon matter and the black spheres (3, 4) to dark matter.}
  \label{modelo}
\end{figure}

Taking an interaction such as \eqref{interacao} as a working hypothesis, we consider this simple mechanical model for a cluster: a system consisting
of two pairs of point masses $m_1$, $m_2$, $m_3$ and $m_4$ connected by two massless poles of length $l_{\text{b}}$ and $l_{\text{d}}$, which
correspond to the length of the semi-major axes of the baryon and dark matter distributions respectively, spinning freely around an axis fixed at the
center of mass, interacting only through Newtonian gravity (see figure \ref{modelo}). Keeping the rotation axis fixed is merely a consequence of
assuming that the dark and baryon components have both zero linear momentum, and not an \emph{a priori} simplification. From angular momentum
conservation, we may assume without loss of generality that the system is confined to the $xy$ plane.

Therefore, the Lagrangian for this model can be written in terms of the angles between the poles and the $x$ axis. Thus,

\begin{multline}\label{lagrangiana}
  \mathcal{L} = \frac{1}{8} \left[ \left( m_1 + m_2 \right) l_{\text{b}}^2 \dot{\theta}_{\text{b}}^2 + \left( m_3 + m_4 \right) l_{\text{d}}^2 \dot{\theta}_{\text{d}}^2 \right] -\\
  - \frac{G}{2} \left[ \frac{m_1 m_3 + m_2 m_4}{D_- (\theta_{\text{b}}, \theta_{\text{d}})} + \frac{m_1 m_4 + m_2 m_3}{D_+ (\theta_{\text{b}},
      \theta_{\text{d}})} \right],
\end{multline}

\noindent
with

\begin{equation}
  D_{\pm} (\theta_{\text{b}}, \theta_{\text{d}}) = \sqrt{l_{\text{b}}^2 + l_{\text{d}}^2 \pm 2 l_{\text{b}} l_{\text{d}} \cos (\theta_{\text{b}} -
    \theta_{\text{d}})}\,.
\end{equation}

The corresponding equations of motion are

\begin{subequations}\label{eqmov}
\begin{gather}
\begin{split}\label{thetab}
  \ddot{\theta}_{\text{b}} = -\frac{\dot{m}_1 + \dot{m}_2}{m_1 + m_2} \dot{\theta}_{\text{b}} + 2 G \frac{l_{\text{d}}}{l_{\text{b}}} \frac{\sin
    (\theta_{\text{b}} - \theta_{\text{d}})}{m_1 + m_2} \times\\
\times \left[ \frac{m_1 m_3 + m_2 m_4}{D_-^3 (\theta_{\text{b}}, \theta_{\text{d}})} - \frac{m_1 m_4 + m_2 m_3}{D_+^3 (\theta_{\text{b}},
      \theta_{\text{d}})} \right],
\end{split}\\
\begin{split}\label{thetad}
  \ddot{\theta}_{\text{d}} = -\frac{\dot{m}_3 + \dot{m}_4}{m_3 + m_4} \dot{\theta}_{\text{d}} - 2 G \frac{l_{\text{b}}}{l_{\text{d}}} \frac{\sin
    (\theta_{\text{b}} - \theta_{\text{d}})}{m_3 + m_4} \times\\
\times \left[ \frac{m_1 m_3 + m_2 m_4}{D_-^3 (\theta_{\text{b}}, \theta_{\text{d}})} - \frac{m_1 m_4
      + m_2 m_3}{D_+^3 (\theta_{\text{b}}, \theta_{\text{d}})} \right].
\end{split}
\end{gather}
\end{subequations}

\section{Simulation parameters and results}

For simplicity, we assume $m_1 = m_2 \equiv m_{\text{b}}$ constant and $m_3 = m_4 \equiv m_{\text{d}} (t)$. The effects of dark matter
interaction shape the function $m_{\text{d}} (t)$ according to the linear model $m_{\text{d}} (t) = m_0 + \lambda t$, where $\lambda
\equiv g H_0$. We have abandoned the global expansion in equation \eqref{interacao} and taken the right-hand side to be proportional only to
$\rho_{\text{DM}}$. We consider an expansion of the solution up to first order.

We have scaled the time parameter in terms of the age of the clusters, such that $t = 0$, the starting time of the simulation, corresponds to
their formation epoch, and $t = 1$, the final time, corresponds to today.

\subsection{Fixed interaction parameter}\label{sec:gfixo}

As a preliminary approach, we have run a simulation of several clusters with a fixed interaction parameter $g$, to verify if there is any consequence
at all to the disposition angles when compared to the non-interacting case where $\dot{m}_{\text{d}} = 0$.

We have fit the parameter $\lambda = m_{\text{d}} (1) - m_0$ to $g = 0.15$ from observation \cite{wang-2008} so that the masses reach $m_{\text{d}} (1)
= 0.857 M_{\text{T}}$ \cite{komatsu-2011}, with $M_{\text{T}} = m_{\text{b}} + m_{\text{d}} (1)$, with an initial mass given by

\begin{equation}
m_{\text{d}} (0) = \frac{m_{\text{d}} (1)}{1 + g H_0}\,,
\end{equation}

\noindent
where we have cast the Hubble constant in terms of the rescaled time.

To ensure that the system has enough initial energy not to undergo gravitational collapse, the initial velocities were chosen such that the tension on
the poles would be zero at the initial time, therefore rendering the pole construction less artificial. For small $\Delta \theta \equiv
(\theta_{\text{b}} - \theta_{\text{d}})$, it may not be possible to satisfy this condition for $\dot{\theta}_{\text{b}} (0)$. In such cases, we have
adopted the same initial value computed for $\dot{\theta}_{\text{d}} (0)$, which means that the system is initially coupled and rotates
synchronously.

We have also cast the mass parameters in terms of ratios with respect to the baryon mass of the cluster, which means that we have taken $m_{\text{b}}
= 1$, and the length parameters as ratios with respect to the typical semi-major axis of the baryon component of the cluster. Considering these
factors in mass, time and length, we have adjusted the gravitational constant $G$ accordingly. The remaining parameters used in our simulation were
therefore $M_{\text{T}} = \nicefrac{20}{3}$, $l_{\text{b}} = 1$ and $l_{\text{d}} = 1.1$. Using these parameters we have integrated the system
\eqref{thetab},\eqref{thetad} using a fourth-order Runge--Kutta method for a gaussian distribution of initial angular differences centered at
0\degree{} with $\sigma = 3.3\degree$ to match the angular distribution provided by simulations through the $\Lambda$CDM-based
predictions \cite{lee-2010}.

\begin{figure}[!htp]
  \centering
  \includegraphics[width=.45\textwidth]{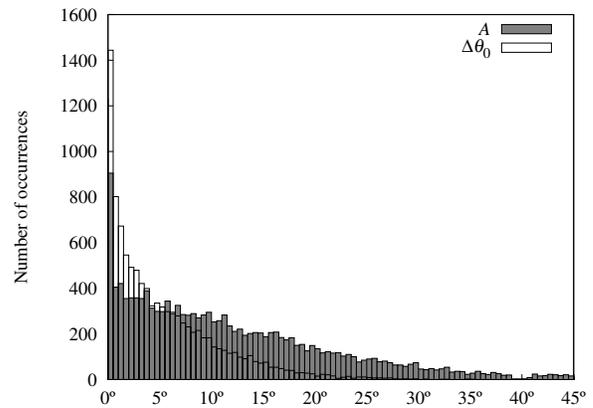}
  \caption{Initial angular differences and final oscillation
    amplitudes for a simulated cluster with parameters given in section \ref{sec:gfixo}, interaction parameter $g = 0.15$ and 20,000 initial angle
    differences. There have been no significant occurrences above 45\degree.}
  \label{hist}
\end{figure}

For an initially synchronous system, as is presumably the case for a standard formation of a cluster from the accretion of both dark and baryon
matter, the angular velocities eventually decouple, and the final angular distance $\Delta \theta (1)$ for a starting angular difference $\Delta
\theta_0$ is roughly described by the fitting

\begin{equation}
  \Delta \theta (t) = A e^{\delta t} \cos (\omega t + \varphi) + \xi\,,
\end{equation}

\noindent
where $A$ is the oscillation amplitude. $\varphi$ is an unimportant phase and $\xi$, also in degrees, is small but non-vanishing. At
the final time $t = 1$ we have observed an increased scatter of angle dispersions (see figure \ref{hist}).

For an initially Gaussian distribution, in the non-interacting case
the final amplitude retains the initial shape, as can be seen by
setting $\dot{m}_{\text{d}} = 0$ in the system \eqref{eqmov}. When the
interaction is present, at the final time the distribution becomes
more scattered while remaining spread around zero, which is in
accordance with the data \cite{lee-2010}.


\subsection{Genetic search for the best-fit interaction parameter}

For more robust comparisons, in order to find a best-fit value for the interaction strength $g$, we need an idea of its correlation with the angular
distribution measured by observations. In order to achieve a figure we have run the program with the interaction turned on and searched for the best
fit in comparison with the observed spread of the clusters.

The search was implemented via a genetic algorithm whose fitness function was the $\chi^2$ computed from the difference between the observational data
and the calculated frequency of the angle $\Delta \theta$ in the simulated clusters. We have used the same initial gaussian sample of angles as in
section \ref{sec:gfixo}. We started with random values of $g$ ranging from 0 to 5. The fitness function was given by the $\chi^2$ between the
simulated frequencies and the observed data from \cite{lee-2010}.

\begin{figure}[!htp]
  \centering
  \includegraphics[width=.45\textwidth]{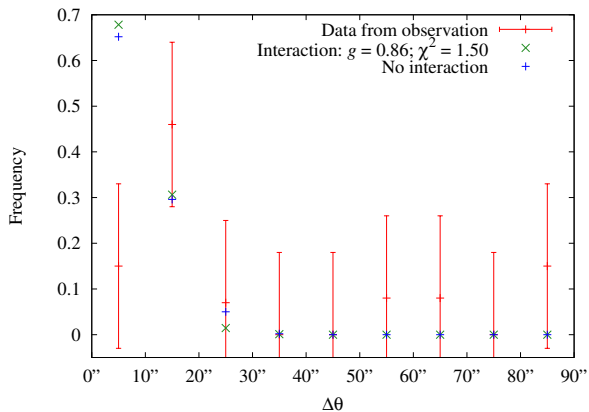}
  \caption{Results of the best-fit search for the interaction parameter for the low uncertainty set of observed clusters from \cite{oguri-2010}, for
    1,000 simulated clusters, in comparison with the non-interacting case.}
  \label{gcompare}
\end{figure}

We can use the raw data from \cite{lee-2010} or the more narrow data selected in that reference. The former leads to an interaction constant
$g=1.75$ which is large when compared with other previous analyses \cite{wang-2007,wang-2008,abdalla-2009,abdalla-2010}. However, taking the more
restricted set of clusters, one finds the result shown in figure \ref{gcompare}, leading to an interaction strength

\begin{equation}
g=0.86
\end{equation}

\noindent
with $\chi^2=1.5$. The above value, although not completely ruling out non-interaction, fits the observed values better than the non-interacting case
(for which the adjustment results in $\chi^2=1.65$. Thus, there is a hint that such observations are compatible with a non vanishing interaction in
the dark sector, a conclusion which points, once more, towards a dark energy different from a simple cosmological constant.

It should be noted that, although we know that this very simplified Classical Mechanical model does not describe the full dynamics of galactic
clusters, the fact that it successfully reproduces (or better approximates) an important feature of the observation data is a hint that dark matter
interaction may be a very basic phenomenon with detectable influences even in the classical realm of interactions.

\section*{Acknowledgements}

This work is supported by FAPESP and CNPq.

\bibliography{referencias}

\end{document}